\documentclass[aps,twocolumn,floats,prd,nofootinbib,superscriptaddress,10pt]{revtex4-1} 

\usepackage[dvips]{graphicx}
\usepackage{graphicx,amsmath,amsfonts,amssymb,slashed}
\usepackage{bbold,wasysym}
\usepackage{graphicx}
\usepackage{here}
\usepackage[usenames,dvipsnames]{xcolor} 
\usepackage{soul}
\usepackage{hyperref}

\definecolor{RedWine}{rgb}{0.743,0,0}
\definecolor{RoyalBlue}{rgb}{0.25,.41,.88}
\setstcolor{Blue}

\renewcommand{\sec}{\ensuremath{\mathrm{s}}}

\newcommand{\km}{\ensuremath{\mathrm{km}}}

\newcommand{\VEV}[1]{\left<{1}\right>}

\newcommand{\Msmbh}{\ensuremath{M_{\mathrm{SMBH}}}}
\newcommand{\Mpbh}{\ensuremath{M_{\mathrm{PBH}}}}
\newcommand{\Rsp}{\ensuremath{R_{\mathrm{sp}}}}
\newcommand{\Rs}{\ensuremath{R_{\mathrm{s}}}}
\newcommand{\rhosp}{\ensuremath{\rho_{\mathrm{sp}}}}
\newcommand{\gammasp}{\ensuremath{\gamma_{\mathrm{sp}}}}

\begin{document}

\title{Primordial-black-hole mergers in dark-matter spikes}

\author{Hiroya Nishikawa}
\affiliation{Department of Physics and Astronomy, Johns Hopkins
     University, 3400 N.\ Charles Street, Baltimore, Maryland 21218, USA}
\author{Ely D.\ Kovetz}
\affiliation{Department of Physics and Astronomy, Johns Hopkins
     University, 3400 N.\ Charles Street, Baltimore, Maryland 21218, USA}
\author{Marc Kamionkowski}
\affiliation{Department of Physics and Astronomy, Johns Hopkins
     University, 3400 N.\ Charles Street, Baltimore, Maryland 21218, USA}
\author{Joseph Silk}
\affiliation{Department of Physics and Astronomy, Johns Hopkins
     University, 3400 N.\ Charles Street, Baltimore, Maryland 21218, USA}
     \affiliation{Institut d'Astrophysique de Paris, UMR 7095, CNRS, UPMC Univ. Paris VI, 98 bis Boulevard Arago, 75014 Paris,
	France}
\affiliation{BIPAC, Department of Physics, University of Oxford, Keble Road, Oxford OX1 3RH, United Kingdom}

\begin{abstract}
It has been suggested that primordial black holes (PBHs) of roughly 30 solar masses could make up the dark matter and if so, might account for the recent detections by LIGO involving binary black holes in this mass range.
It has also been argued that the super-massive black holes (SMBHs) that reside at galactic centers may be surrounded by extremely-dense dark-matter (DM) spikes.
Here we show that the rate for PBH mergers in these spikes may well exceed the merger rate considered before in galactic dark-matter halos, depending on the magnitudes of two competing effects on the DM spikes: depletion of PBHs due to relaxation and replenishment due to PBHs in loss cone.
This may provide a plausible explanation for the current rate of detection of mergers of 30-solar-mass black holes, even if PBHs make up a subdominant contribution to the dark matter.
The gravitational-wave signals from such events will always originate in galactic centers, as opposed to those from halos, which are expected to have little correlation with luminous-galaxy positions.
\end{abstract}

\maketitle

\section{Introduction}
Over the first few months of coincident measurement, the LIGO interferometers have detected gravitational waves from several mergers of black hole binaries~\cite{Abbott:2016blz,Abbott:2016nmj,Abbott:2017vtc}. 
Two of these events involved the mergers of black holes with masses estimated to be near 30 $M_\odot$.
While these may simply be the endpoints of massive stars\footnote{The current detections are consistent with a simple stellar-black-hole~\cite{Belczynski:2009xy,Belczynski:2016obo} power-law mass function with a high mass cutoff (see Refs.~\cite{TheLIGOScientific:2016pea,Kovetz:2016kpi} and supplemental materials of Ref.~\cite{Abbott:2017vtc}).}, an alternative explanation that is tempting to consider is that these are primordial black holes (PBHs)~\cite{Zel'dovich:1967,Carr:1974nx,Chapline:1975,Frampton:2010sw}, which are formed deep in the radiation-dominated era.
This idea is especially intriguing as there remains the possibility that such PBHs could account for the dark matter (DM) in the Universe~\cite{Bird:2016dcv,Clesse:2016vqa,Sasaki:2016jop,Kashlinsky:2016sdv,Blinnikov:2016bxu,Dolgov:1992pu,Dolgov:2008wu}.
Although tensions between 30 $M_\odot$ PBH dark matter and various astrophysical observations have been discussed~\cite{Alcock:1996yv,Tisserand:2006zx,Mediavilla:2017bok,Ricotti:2007au,Ali-Haimoud:2016mbv,Brandt:2016aco,Raidal:2017mfl,Poulin:2017bwe} (see~\cite{Munoz:2016tmg,Schutz:2016khr,Kovetz:2017rvv,Venumadhav:2017pps,Diego:2017drh,Clesse:2016ajp} for promising future constraints), each comes with some caveats.  Given the fundamental nature of the dark matter, as well as any obvious solution, continued attention to the possibility of 30 $M_\odot$ PBH dark matter is still warranted.

The validity of the PBH scenario for LIGO's more massive events depends strongly on the predicted rates of their binary formation and merger, which are difficult to determine.
For example, the probability that early-formed binaries remain undisrupted until they merge in the local volume detectable by LIGO is uncertain~\cite{Sasaki:2016jop,Hayasaki:2009ug,Ali-Haimoud:2017rtz}.
Meanwhile, the properties of the smallest dark-matter halos, where two-body binary formation through emission of gravitational waves in close encounters is most efficient~\cite{Bird:2016dcv}, can only be estimated based on extrapolations which cannot be corroborated directly by observations.
It is therefore worthwhile to investigate other channels of PBH binary formation.  

In this work, we set out to calculate the overall rate at which PBH mergers occur in the vicinity of super massive black holes (SMBH), where we may expect a significant enhancement due to the higher DM density.
In particular, Gondolo and Silk~\cite{Gondolo:1999ef} suggested that an extremely dense DM spike could form near a galactic center if galactic halos are cusped, as favored in N-body simulations of galaxy formation~\cite{Graham:2005xx,Stadel:2008pn,Navarro:2008kc}.
Given the uncertainty in the DM-spike profile (see Ref.~\cite{Ullio:2001fb}, for example), and as we cannot observe or simulate decisively the innermost regions around central SMBHs, our aim is merely to derive an estimate of the resulting PBH merger rate at the order-of-magnitude level, to be compared with the current LIGO estimate for $\sim30\,M_\odot$ black holes: $0.5-12\, {\rm Gpc^{-3} yr^{-1}}$ at $90\%$ confidence~\cite{TheLIGOScientific:2016pea,Abbott:2017vtc}.
We present the GW-detection rate in two limits of the DM-spike evolution, in order to account for the effects of two competing mechanisms: PBH depletion due to relaxation and PBH replenishment through loss-cone refilling.
Our results demonstrate that the PBH mergers in the originally proposed DM spikes may generate a significant contribution to the detection rate inferred by LIGO, while relaxation due to two-body interactions between PBHs within the DM spikes may suppress the rate down to as small as its $\sim 1 \%$ in the absence of efficient PBH-replenishment processes.
Due to the lack of our understanding of the exact properties of the DM-density profile in the proximity of a central SMBH, the mass function of SMBHs in the Universe, and the effect of the PBH replenishment, we conclude that the actual rate lies between the two results, depending largely on the magnitudes of the two competing effects.
This may render the GW-emission rate from galactic centers comparable to the LIGO-inferred rate.

Our paper is laid out as follows: In Sec.~\ref{sec:Model} we present our model for estimating the rate of PBH mergers in DM spikes.
We first express the spike profile as a function of the SMBH mass, using the mass-dispersion relation and the mass-concentration relation in an Navarro-Frenk-White (NFW) halo profile, and then derive the overall PBH merger rate, 
which also depends on the SMBH mass function.
In Sec.~\ref{sec:Empirics}, we describe our parameter choices for the various empirical relations used in our calculation, comparing different prescriptions adopted from the literature.
The evolution of the DM spikes, mainly due to the effects of gravitational relaxation and replenishment through loss-cone refilling, is then considered in Sec.~\ref{sec:Evolution}.
We show our results and conclusions in Secs.~\ref{sec:Results} and~\ref{sec:Conclusions}, respectively.

\section{Model}\label{sec:Model}
\subsection{The dark-matter spike profile}\label{ssec:Spike_profile}
We consider a SMBH of mass $\Msmbh$ residing in a DM halo which initially has a density profile near the galactic center of the form $\rho(r) \simeq \rho_0 {(r_0/r)}^{\gamma}$, where $\gamma$ is the power-law index and $\rho_0$ and $r_0$  are halo parameters.
As shown in Ref.~\cite{Gondolo:1999ef}, this will lead to the formation of a DM spike of radius $\Rsp(\gamma, \Msmbh) = \alpha_\gamma r_0 {(\Msmbh/(\rho_0 r_0^3))}^{1/(3-\gamma)}$, where the normalization $\alpha_\gamma$ is numerically derived for each power-law index $\gamma$.
The density profile in this spike for $r$ in the range $4\Rs<r<\Rsp$ is given by
\begin{equation}
	\rhosp(r) =\rho_R {\left(1  - \frac{4\Rs}{r} \right)}^3 {\left(\frac{\Rsp}{r}\right)}^{\gammasp},
    \label{eq:rhosp}
\end{equation}
where $\rho_R=\rho_0 {(\Rsp/r_0)}^{-\gamma}$, $\gammasp = (9-2\gamma)/(4-\gamma)$, and $\Rs =2 G \Msmbh/c^2 \simeq 2.95 \ (\Msmbh/\, { M}_\odot)\,{\rm km}$ is the Schwarzschild radius of the SMBH\@. 
Note that the spike density is enhanced in the case of a Kerr black hole, where the spike continues into about  twice the horizon scale~\cite{Ferrer:2017xwm}.

\begin{figure}[t]
        \includegraphics[width=\columnwidth]{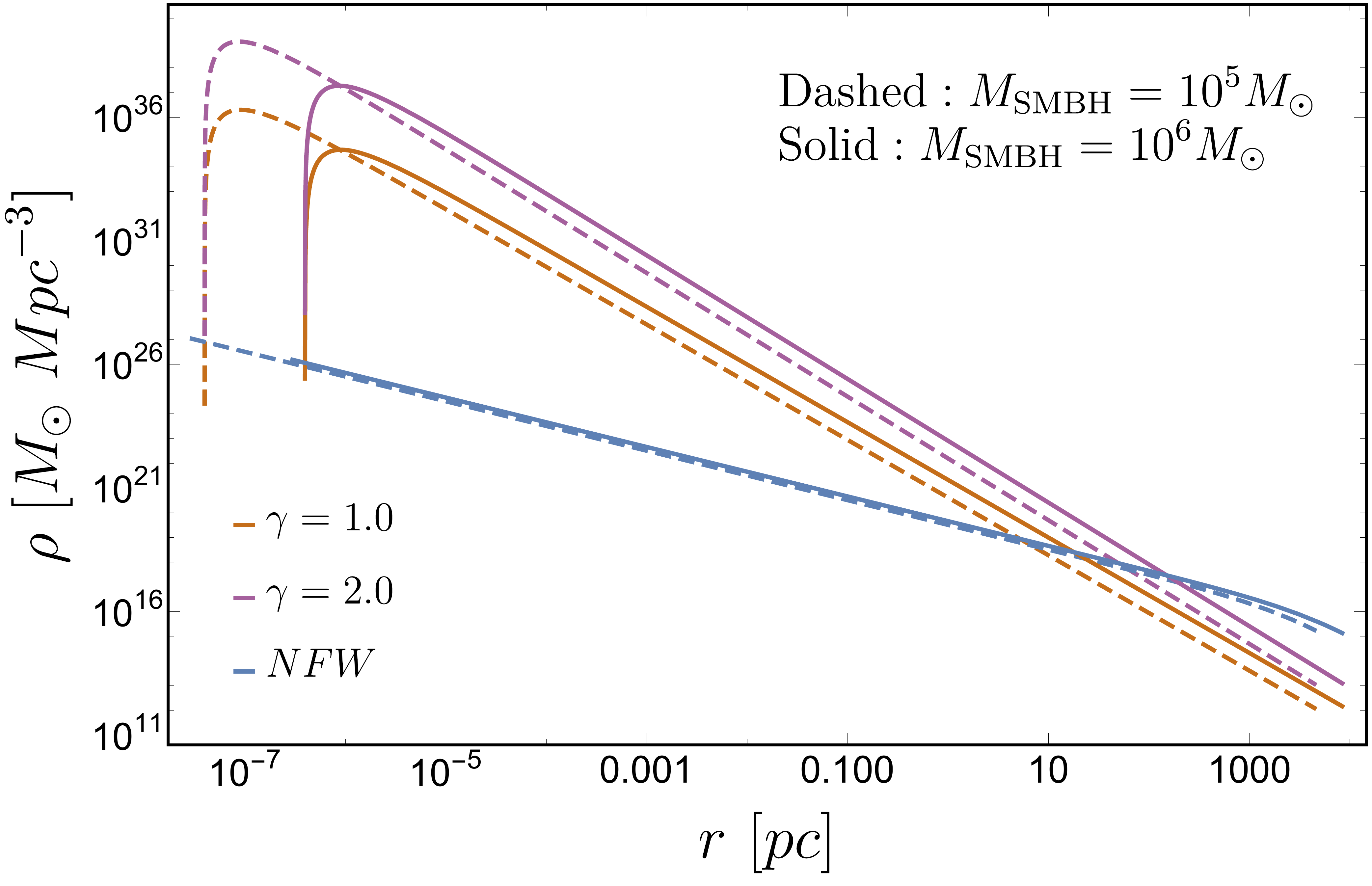}
        \caption{A comparison of the NFW  and spike-density profiles with $\gamma=1$ or $\gamma=2$  surrounding a SMBH 
        with mass $\Msmbh=10^5\, M_{\odot}$ or $\Msmbh=10^6\, M_{\odot}$.
        The spike profiles cross the NFW profile at $r= \Rsp(\gamma,  \Msmbh)$, defining the radius within which the merger rate will be calculated. 
        Note that the NFW profile displays a $\sim r^{-1}$ behavior inside this region, as expected.}\label{fig:rhovsr}
\end{figure}

In Fig.~\ref{fig:rhovsr} we show how $\rhosp(\gamma,r)$ differs from the NFW density profile for $\gamma=1$ and 
$\gamma=2$ (more on the value of $\gamma$ in Sec.~\ref{sec:Empirics}) and $\Msmbh=10^5\, M_{\odot}$ or $\Msmbh=10^6\, M_{\odot}$. 
We see that the density is enhanced by several orders of magnitude in the spike region, and it is therefore worth investigating whether this could have a significant effect on the PBH merger rate.
We note that the evolution of the DM spikes is considered in Sec.~\ref{sec:Evolution}.

\subsection{Relating the SMBH mass to the halo parameters}
We wish to obtain an estimate of the merger rate per halo as a function of the SMBH mass.
In order to relate the halo parameters $\rho_0$ and $r_0$ to $\Msmbh$, we use the $M_{\rm SMBH}$-$\sigma$ relation~\cite{Ferrarese:2000se,Gebhardt:2000fk},
\begin{equation} \label{eq:Msigma}
    \log_{10} (\Msmbh/M_\odot) = a + b\,\log_{10} (\sigma/ 200\, {\km}~{\sec}^{-1}),
\end{equation}
where $a$ and $b$ are empirically determined parameters (see Sec.~\ref{sec:Empirics}). 
The $M_{\rm SMBH}-\sigma$ correlation has a lower scatter than other similar relations, such as the $M_{\rm SMBH}$-$L_{\rm bulge}$ and $M_{\rm SMBH}$-$M_{\rm bulge}$ relations~\cite{Kormendy:1995er,Magorrian:1997hw,Marconi:2003hj}, and is especially useful for our purposes as it relates $\Msmbh$ to the velocity dispersion $\sigma$ of the dark-matter halo, which can be expressed using $\rho_0$ and $r_0$.
We relate $M_{\rm SMBH}$ to $\rho_0$ and $r_0$ by assuming that the dark-matter density profile in the region $r \gg \Rsp$ external to the spike is described by an NFW profile, extending out to the virial radius $r_{\rm vir} > r_0$.
The total mass enclosed within a sphere of radius $r$ is then given by
\begin{equation}
    M(r)= 4\pi \, \rho_0 r_0 \int_{0}^{r} \frac{r \, dr}{{(1+r/r_0)}^2}
        = 4 \pi \, \rho_0 r_0^3 \, g(r/r_0),
        \label{eq:Mr}
\end{equation}
where $g(x)=\log{(1+x)}-x/(1+x)$ [note that we can safely ignore the contribution from the mass of the density spike itself as well as the SMBH at the center since they are negligible compared to $M(r)$].
The circular velocity ${(GM(r)/r)}^{1/2}$, which is maximized at a distance $r/r_0=2.16 \equiv c_{\rm m}$ from the center of an NFW halo, is equal to the one-dimensional halo velocity dispersion $\sigma$, 
\begin{equation} \label{eq:sigma}
    \sigma ^2=\frac{GM(c_{\rm m} r_0)}{c_{\rm m} r_0}
    =\frac{4 \pi G \, \rho_0 r_0^2 \, g(c_{\rm m})}{c_{\rm m}}.
\end{equation}
We now define the halo concentration parameter as $c(M_{\rm vir}) \equiv r_{\rm vir}/r_{0}$, where $M_{\rm vir}=M(r_{\rm vir})$ is the mass enclosed within the virial radius $r_{\rm vir}$. 
Using Eq.~\eqref{eq:Mr}, we then see that
\begin{equation} \label{Mvir1}
    M_{\rm vir}\equiv200 \rho_{\rm crit}\,\left(\frac{4 \pi \, {(c(M_{\rm vir}) r_0)}^3}{3} \right)=
    4 \pi \rho_{0} r_0^3 g(c(M_{\rm vir})),
\end{equation}
which allows us to relate $\rho_0$ and $r_0$ to $\Msmbh$ through Eq.~\eqref{eq:Msigma} and Eq.~\eqref{eq:sigma}.

\subsection{The PBH merger rate}
The PBH merger rate $N_{\rm sp}$ per year in a spike around a SMBH of mass $\Msmbh$ can be calculated using
\begin{equation} \label{eq:N}
   N_{\rm sp}
   = \int_{4 \Rs}^{\Rsp} \frac12 {\left(\frac{\rhosp(r)}{\Mpbh} \right)}^2 \sigma_{m}(r)  v(r) \, d^3r,
\end{equation}
where the merger cross section $\sigma_{m}(r)$ is given by~\cite{Bird:2016dcv}
\begin{equation}
    \sigma_{m}(r)
        = 1.4 \times 10^{-14}\, {\left(\frac{\Mpbh}{30\,
        M_\odot} \right)}^2 {\left(\frac{v(r)}{200~{\rm km}~{\sec}^{-1}}
        \right)}^{-\frac{18}{7}}{\mathrm{pc}^2}. 
\end{equation}
For the relative velocity we use the circular speed $v(r) = {(G M_{\rm SMBH}/r)}^{1/2}$ at each radius $4\Rs < r < \Rsp$, since the total mass inside the spike is negligible compared to the mass of the central SMBH\@. 

\begin{figure}[t]
    \includegraphics[width=\columnwidth]{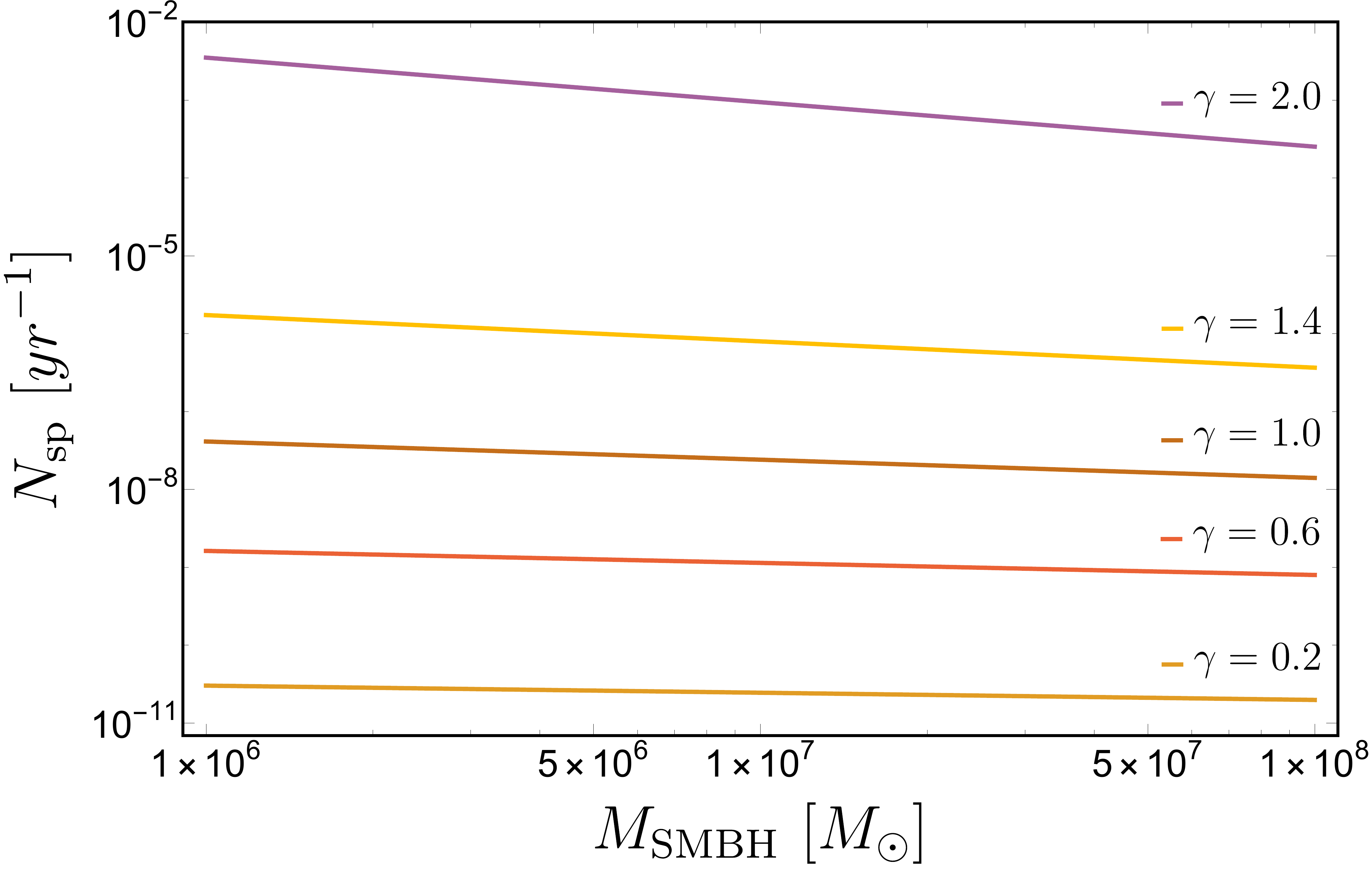}
    \caption{The PBH merger rate $N_{\rm sp}$ from the original DM spikes per year per halo as a function of the SMBH mass $\Msmbh$.
    We see that the more massive the SMBH at the center, fewer PBH mergers happen, which is opposite to the behavior observed in Ref.~\cite{Bird:2016dcv}.}\label{fig:Nspvsm2}
\end{figure}

In Fig.~\ref{fig:Nspvsm2}, we plot $N_{\rm sp}$ as a function of $M_{\rm SMBH}$ for several values of spike power-law index $\gamma$.
Interestingly, the rate does not depend on the PBH mass $M_{\rm PBH}$, since the effect of decreased abundance with larger $M_{\rm SMBH}$ is compensated by the increase in the cross section.

In order to calculate the overall contribution to the PBH merger rate from density spikes around SMBHs, the final step is to convolve the merger rate per SMBH mass, Eq.~\eqref{eq:N}, with a mass function $\phi(M_{\rm SMBH})$ of SMBHs to obtain the total rate per unit volume, 
\begin{equation}\label{eq:Gammaint}
    \Gamma = \int_{M_{\min}}^{M_{\max}} \, N_{\rm sp}(\Msmbh) \, \phi(\Msmbh) \, d\Msmbh.
\end{equation}
This quantity is implicitly dependent on the parameters and functional relations introduced in the derivation above, 
\begin{equation}\label{eq:Gamma}
    \Gamma=\Gamma(\gamma, a, b, c(M_{\rm vir}),\phi(\Msmbh),M_{\min},M_{\max}),
\end{equation}
stemming from the steepness of the density profile, the $M_{\rm SMBH}-\sigma$ relation, the halo concentration, the SMBH mass function and the minimum and maximum SMBH masses. 
In the next section, we explore the relevant ranges for these parameters, based on empirical fits to various datasets, and motivate the choices we make in our final calculation.

\section{Empirics}\label{sec:Empirics}
\subsection{Density profile}
Evidence from numerical simulations~\cite{Graham:2005xx,Stadel:2008pn,Navarro:2008kc} and some analytic arguments suggest that the density profile has a power-law dependence on the radius at small radius. 
While there are some reasons to believe that the power-law index is $\gamma\simeq1$, as seen in NFW~\cite{Navarro:1995iw} and Einasto~\cite{Graham:2005xx} profile, there are also arguments that it may take on other values.
We thus explore in our analysis below values $0<\gamma <2$.

\subsection{The \texorpdfstring{$M_{\rm SMBH}-\sigma$}{} relation}
We follow Ref.~\cite{Gultekin:2009qn} which finds $a=8.12 \pm 0.08$ and $b=4.24 \pm 0.41$ to be a good fit for all types of galaxies, and for a comparison consider the results of Ref.~\cite{Kormendy:2013dxa} as well.
We found that these uncertainties give rise to at most a $10\%$ error in our final results, much smaller than that induced by the other factors considered in Sec.~\ref{ssec:SMBH_mass_function}.

\subsection{The mass-concentration relation}
We use the mass-concentration relation found in Ref.~\cite{Prada:2011jf}, which is based on a fit to multiple $\Lambda$CDM N-body simulations, in which $c$ is expressed as a function of redshift $z$ and $M_{\rm vir}$.
In our calculation, we set $z=0$ since the redshifts ($z\leq0.3)$ detectable by LIGO are relatively low.
The errors in this best-fit approximation are less than a few percent (see Fig.~10 in Ref.~\cite{Prada:2011jf}), and they translate into roughly the same percent error in our final results.
Even if we assume the error in the mass-concentration relation to be much larger, it would not shift our final result nearly as much as the other factors considered below.

\subsection{The SMBH mass function}\label{ssec:SMBH_mass_function}
Compared to the parameters and the functional relation already discussed, the SMBH mass function $\phi(\Msmbh)$ turns out to generate substantial uncertainty in the final rate $\Gamma$ both for the original DM spikes and the (fully) relaxed DM spikes considered in Sec.~\ref{sec:Evolution}, since the current estimate of the mass function is highly uncertain.
To account for this uncertainty, we will compare three different empirical SMBH mass functions. 
In Ref.~\cite{Dzanovic:2006px}, a sample of $\sim9000$ SDSS galaxies was used to infer the spheroid and disk galaxy luminosity functions, and based on the assumption that all spheroids contain SMBHs at their center, the SMBH mass function is derived to be,
\begin{equation}
    \phi(\Msmbh)= 10^9 \left(\frac{\phi_0}{M_{\ast}}\right) {\left( \frac{\Msmbh}{M_{\ast}} \right)}^\alpha 
    e^{-{\left({\Msmbh}/{M_{\ast}} \right)}^\beta}, 
\end{equation}
with $\phi_0 = 0.0029\,h^3\, {\rm Mpc}^{-3}$, $\alpha=-0.65$, $M_{\ast}=4.07\times10^7\,h^{-2}\, M_\odot$ and $\beta=0.6$.
Ref.~\cite{Vika:2009ef} performed a  similar analysis based on the same spheroid-luminosity to SMBH-mass relation, using $1743$ galaxies from the Millennium Galaxy Catalogue~\cite{Liske:2002nk}, finding
\begin{equation}
\phi(M_{\rm SMBH}) = \phi_{\ast} {\left(\frac{M_{\rm SMBH}}{M_{\ast}}
\right)}^{\alpha+1} \exp\left[1-\left(\frac{M_{\rm SMBH}}{M_{\ast}} 
\right) \right],
\end{equation}
with best-fit values $\log\phi_{\ast}=-3.15$, $\log{M_{\ast}/M_\odot}=8.71$ and $\alpha=1.20$.
Meanwhile, Ref.~\cite{Shankar:2004ir} used kinematic and photometric data to estimate the SMBH mass function based on the empirical relation between the halo velocity dispersion and the SMBH mass, resulting in
\begin{equation}
	\phi(M_{\rm SMBH})
    = \phi_{\ast}{\left(\frac{M_{\rm SMBH}}{M_{\ast}}\right)}^{\alpha+1}
    \exp\left[-{\left(\frac{M_{\rm SMBH}}{M_{\ast}}\right)}^{\beta}\right],
\end{equation}
with best-fit values $\phi_{\ast} = 7.7 \times 10^{-3} \mathrm{Mpc}^{-3}$, $M_{\ast}=6.4 \times 10^7 M_{\odot}$, $\alpha = -1.11$ and $\beta = 0.49$.

\begin{figure}[t]
    \begin{center}
        \includegraphics[width=\columnwidth]{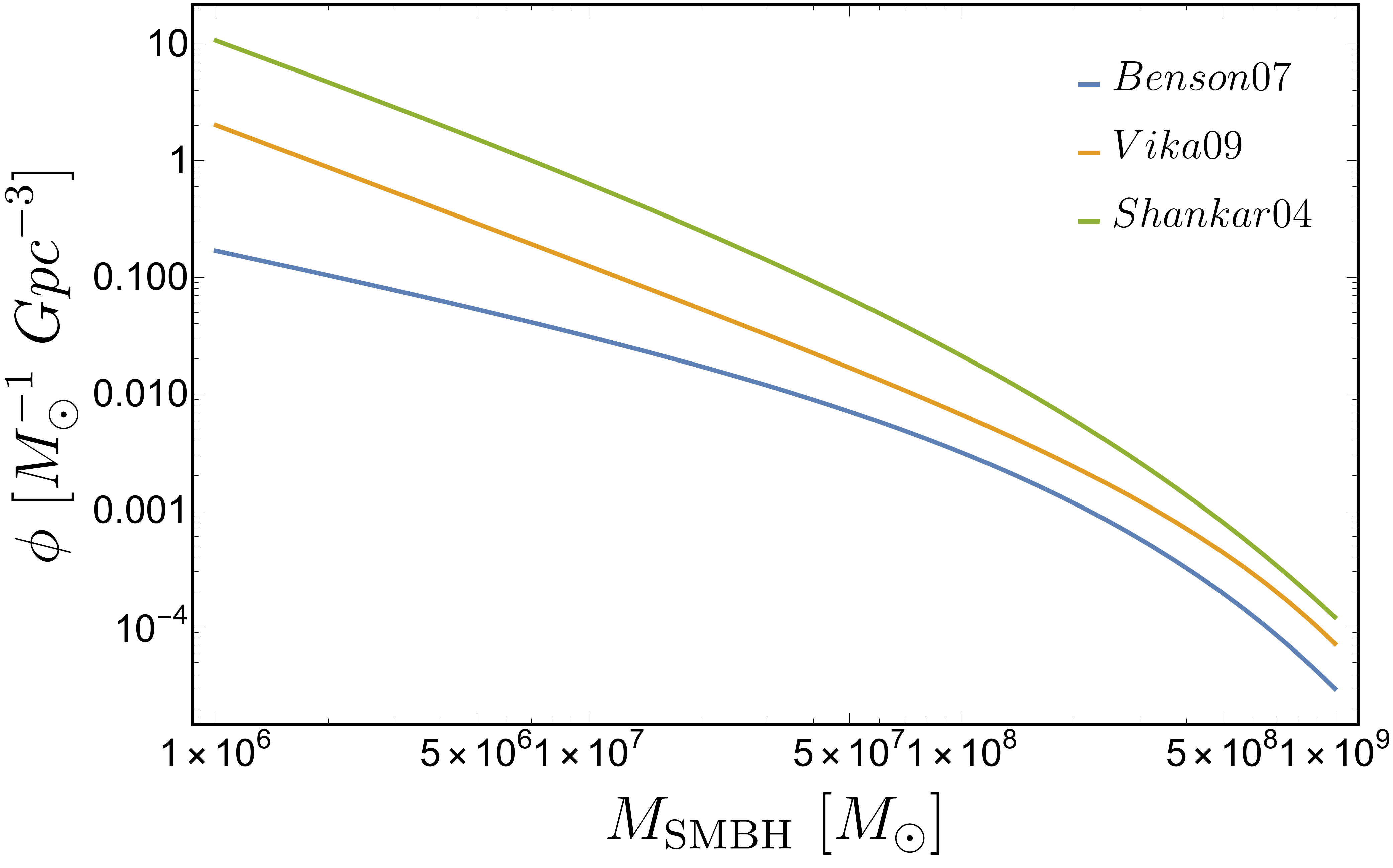}
        \caption{Three different SMBH mass functions $\phi(\Msmbh)$: Benson07~\cite{Dzanovic:2006px}, Vika09~\cite{Vika:2009ef} and Shankar04~\cite{Shankar:2004ir}.
        All three curves fall rapidly as $\Msmbh$ increases.
        In all cases $\phi(\Msmbh)$ peaks towards the lower cut off $M_{\min}$, where there is more than an order-of-magnitude difference.}\label{fig:bhmfvsm}
    \end{center}
\end{figure}

There has been recent interest in the possibility that dwarf galaxies and even globular clusters contain SMBH, based on observational~\cite{Baldassare:2016cox} and even theoretical indications~\cite{Silk:2017yai}.
This can only augment the final rate $\Gamma$, but we will subsequently not consider this possibility. 
Here we assume that $\Msmbh$ in galactic centers ranges from approximately $M_{\min} = 10^5 - 10^6 \, M_\odot$ to $M_{\max} =10^9 - 10^{10} \, M_\odot$, where the three fits above are typically valid, since $M_{\min}$ and $M_{\max}$ introduce uncertainties in the rate $\Gamma$ for the original DM spikes and the relaxed DM spikes, respectively, as shown in Sec.~\ref{sec:Results}.
In Fig.~\ref{fig:bhmfvsm} we plot the three mass functions above for comparison.
As can be seen, lower mass SMBHs are evidently far more abundant in the Universe, and therefore will contribute significantly to our final rate $\Gamma$ for the original DM spikes.
This also means that the choice of $M_{\max}$ does not significantly affect the result (see Fig.~\ref{fig:ratevsgamma}).
Consequently, for the original DM spikes, to appreciate the uncertainties in both the SMBH mass function and the lower cutoff $M_{\min}$, we apply the three mass functions in Fig.~\ref{fig:bhmfvsm} separately and compare between two different lower mass cutoffs, $M_{\min}=10^5\, M_\odot$ and $M_{\min}=10^6\, M_\odot$.
We note that, however, as the dependence on $M_{\max}$ is more important for the relaxed DM spikes, in Fig.~\ref{fig:raterelvsgamma} we compare the results with two different upper mass cutoffs instead, $M_{\max}=10^9\, M_\odot$ and $M_{\max}=10^{10}\, M_{\odot}$.

\section{Spike evolution}\label{sec:Evolution}
We consider the two competing effects on the DM spikes: relaxation due to two-body interactions among the comprising PBHs and replenishment of PBHs due to loss-cone refilling.
We first describe the prescription for quantitatively deriving the relaxed DM spikes in the limit of negligible replenishment effect (i.e.\ the relaxed-spike limit).
The rate from the relaxed DM spikes provides the lower bound for the actual GW-detection rate since such DM-density profiles are reached in the absence of efficient PBH repopulation mechanisms.

In order to estimate the effect of relaxation in this limit, we first find the radius $R_{\mathrm{core}}$ at which the relaxation time $t_{\mathrm{relax}}$ becomes compatible to Hubble time $t_{\mathrm{H}}$.
The local relaxation time $t_{\mathrm{relax}}$ is given by~\cite{Sellwood:2010qa}
\begin{equation}\label{eq:relaxation}
    t_{\mathrm{relax}} = \frac{{v(r)}^3}{8 \pi G^2 \, 30 m_{\odot} \, \rhosp(r) \, \log{(b_{\max}/b_{\min})}},
\end{equation}
and $R_{\mathrm{core}}$ is obtained by solving $t_{\mathrm{relax}} = t_{\mathrm{H}}$.
The impact parameters $b_{\min}$ and $b_{\max}$ are the Schwarzschild radius $\Rs$ of the PBHs and the characteristic size of the gravitational system in question -- $\Rsp$ for the DM spike structure, respectively.
Note that their exact values are not important since the dependence is only logarithmic, and $\log{(b_{\max}/b_{\min})}$ stays $\sim 20$.

We assume that a core forms in the inner region of the DM spike $r \leq R_{\mathrm{core}}$ where $t_{\rm relax}$ is less than Hubble time.
For example, when $\Msmbh = 10^7  \, M_{\odot}$ and $\gamma = 1$, $R_{\mathrm{core}} = 0.23 \Rsp$.
This implies that while the majority of the DM spike remains undisturbed, the more-enhanced part ($r \ll \Rsp$) smooths into a core, leading to a subdued contribution to the merger rate $N_{\rm sp}$ from the innermost region.
This suppression becomes more significant for the DM spikes with larger $\gamma$ and smaller $\Msmbh$, as they lead to more density enhancement in the innermost region [see Eq.~\eqref{eq:rhosp} and Fig.~\ref{fig:rhovsr}].
In fact, the entire regions of the DM spikes shown in Fig.~\ref{fig:rhovsr} would undergo relaxation and form cores in less than one Hubble time.

Finally, we assume that the excess mass is distributed to the outskirt of the spike $R_{\mathrm{core}} \leq r \leq \tilde{R}_{\rm sp}$ as a core forms.
$\tilde{R}_{\rm sp}$ is determined numerically so that the total mass of the initial DM spike equates the total mass of the relaxed DM spike $\tilde{\rho}_{\rm sp}$ in the core and the extended region.
\begin{align}
    \tilde{\rho}_{\rm sp} (r)&=
    \begin{cases}
        \rhosp(R_{\mathrm{core}}) & (4\Rs<r<R_{\mathrm{core}}) \\
        \rhosp(r) & (R_{\mathrm{core}}<r<\tilde{R}_{\rm sp})
    \end{cases}
\end{align}
The upper limit $\Rsp$ of the integral in Eq.~\eqref{eq:N} is then set to $\tilde{R}_{\rm sp}$ to include the contribution from this region.
The result of such DM density profiles is shown in Fig.~\ref{fig:raterelvsgamma}.

We turn to the effect of PBHs on loss-cone orbits that plunge into the DM spikes and may contribute to the merger rate.
Loss cone, originally studied in the context of understanding the evolution of massive BHs at the center of globular clusters~\cite{Lightman:1977zz, Frank:1976uy}, refers to the ensemble of orbits that would experience tidal disruption or direct capture by a central SMBH\@.
The number of stellar objects that lie within loss cone, and how they are re-supplied once removed after one periapsis passage, determine the flux of mass into the central region.
Repopulation of such orbits is facilitated by gravitational encounters between the stellar objects for spherical halos with collisionally-relaxed region (nuclei), while non-spherical (axisymmetric or triaxial) cases also allow more efficient noncollisional feeding onto the orbits (see Ref.~\cite{Merritt:2013awa} for a comprehensive review).

Recent study~\cite{Vasiliev:2013az} estimated the rate of such captures for realistic galaxies possessing the initial-density profile $\rho(r) \simeq \rho_0 {(r_0/r)}^{\gamma}$ with $0.5 \leq \gamma \leq 2.0$, same as we stipulated in Sec.~\ref{ssec:Spike_profile}, for radius roughly corresponding to the outside of the DM-spike region ($\gtrsim 10 \, \rm pc$).
They showed the stationary capture rate of $10^{-4} - 10^{-6} M_{\odot} \rm yr^{-1}$ for $M_{\rm SMBH} = 10^6 - 10^{10} M_{\odot}$ in spherically symmetric case.
While it is not certain how effectively such captures of PBHs may lead to an increase in the number of PBHs within the DM-spike region, we argue that this process may as well affect the evolution of the DM spikes; if the efficiency of replenishment becomes compatible to the rate of PBH depletion due to relaxation within the DM spikes, the decrease in density may not be as significant as that of the relaxed-spike limit.
Since the capture rate becomes higher for smaller halos~\cite{Vasiliev:2013az}, we note that this consideration becomes more important to such halos, whose contribution to the total rate $\Gamma$ gets enhanced once convolved with the mass function (as discussed in Sec.~\ref{ssec:SMBH_mass_function}).
In addition, as the capture rate only increases in nonspherically symmetric cases~\cite{2017MNRAS.464.2301G}, and as whether such high rates can be reached depends strongly on the initial conditions such as the orbital distribution of stellar objects~\cite{Merritt:2013awa}, the significance of the replenishment process is highly dependent on the specifics of halos we consider.
We note that the formation of nonspherical DM distribution around galactic centers (e.g.~\cite{Naoz:2014zda}) may further boost the binary PBH-merger rate from such regions, as more binaries in highly eccentric orbits can be attained through an enhancement in Lidov-Kozai process~\cite{Petrovich:2017otm}.
Lastly, since such considerations are also important for estimating the LISA detection rate of GW signals from stellar compact objects inspiraling around a central SMBH~\cite{Sigurdsson:2003ei}, determining the LISA detection rate of stochastic-GW signals from PBHs orbiting in the DM spikes~\cite{Kuhnel:2018mlr} may also require further analyses of the replenishment process (as well as the depletion process).
As such, we leave quantitative study of this effect to future work, and compute the final rate in Sec.~\ref{sec:Results} based on the DM-density profile in the two limits: the original DM spikes (i.e.\ the original-spike limit), and the (fully) relaxed DM spikes.
The former corresponds to the limit when the PBH-replenishment process is highly efficient, counteracting the effect of relaxation within the DM spikes, and the latter corresponds to the limit of negligible PBH refilling.

\section{Results}\label{sec:Results}
\begin{figure}[t]
        \includegraphics[width=\columnwidth]{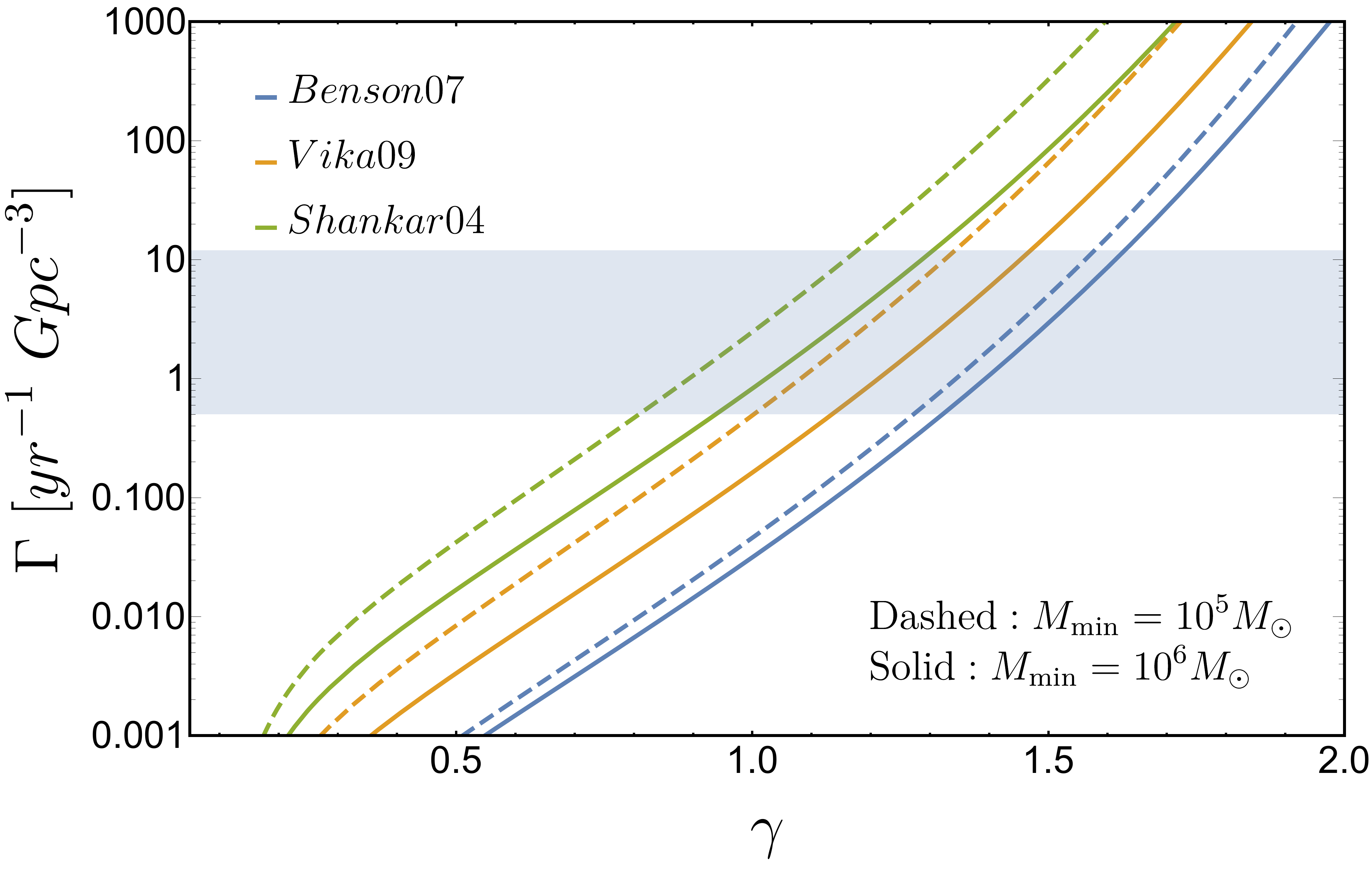}
        \caption{$\Gamma(\gamma)$ from the original DM spikes with 3 different halo mass functions.
        The result of the relaxed DM spikes is shown in Fig.~\ref{fig:raterelvsgamma}.
        The shaded region represents the rate $0.5 \sim 12\, { \rm Gpc^{-3} yr^{-1}}$ estimated by LIGO~\cite{TheLIGOScientific:2016pea}.
        Each color represents each halo mass function corresponding to the Fig.~\ref{fig:bhmfvsm}.
        We plot $\Gamma$ with $M_{\min}=10^5 \, M_{\odot}$ as dashed lines and $M_{\min}=10^6 \, M_{\odot}$ as solid lines.
        }\label{fig:ratevsgamma}
\end{figure}

We show the final rate for the two limiting cases, the rate from the original DM spikes and the rate from the relaxed DM spikes.
Based on the parameter choices described in Sec.~\ref{sec:Empirics}, we proceed to calculate our final results for the overall merger rate $\Gamma(\gamma)$.
As for the original DM spikes, $\gamma$, $\phi(\Msmbh)$ and $M_{\min}$ are found to dominate the final rate $\Gamma \simeq \Gamma (\gamma,\phi(\Msmbh),M_{\min})$, rather than the rest of the factors that appear in Eq.~\eqref{eq:Gamma}.
Using the mass function prescriptions described in Sec.~\ref{sec:Empirics}, and setting the minimum mass $M_{\min}$ to either $10^5 \, M_\odot$ or $10^6 \,  M_\odot$ and the maximum mass $M_{\max}$ to $10^{9}\, M_\odot$, we show in Fig.~\ref{fig:ratevsgamma} the rate $\Gamma$ from the original DM spikes as a function of $\gamma$.
Figure~\ref{fig:ratevsgamma} includes a band indicating the $90\%$ confidence interval for the merger rate of black-hole binaries (similar to GW150914)~\cite{TheLIGOScientific:2016pea}. 
We see that the contribution to the PBH merger rate from the original DM spikes alone can produce the full rate inferred by LIGO, depending strongly on the value of $\gamma$ and also on the adopted mass function.
For $\gamma\sim1$, the rate ranges from $1\, { \rm Gpc^{-3} yr^{-1}}$---consistent with the contribution from all halos outside their spike region~\cite{Bird:2016dcv}---down to roughly $10\%$ of that.
As this rate may be achieved when the smoothing of the DM spikes due to relaxation is counteracted by the replenishment of PBHs, we show this rate as the upper bound for the GW-detection rate from galactic centers.

\begin{figure}[t]
        \includegraphics[width=\columnwidth]{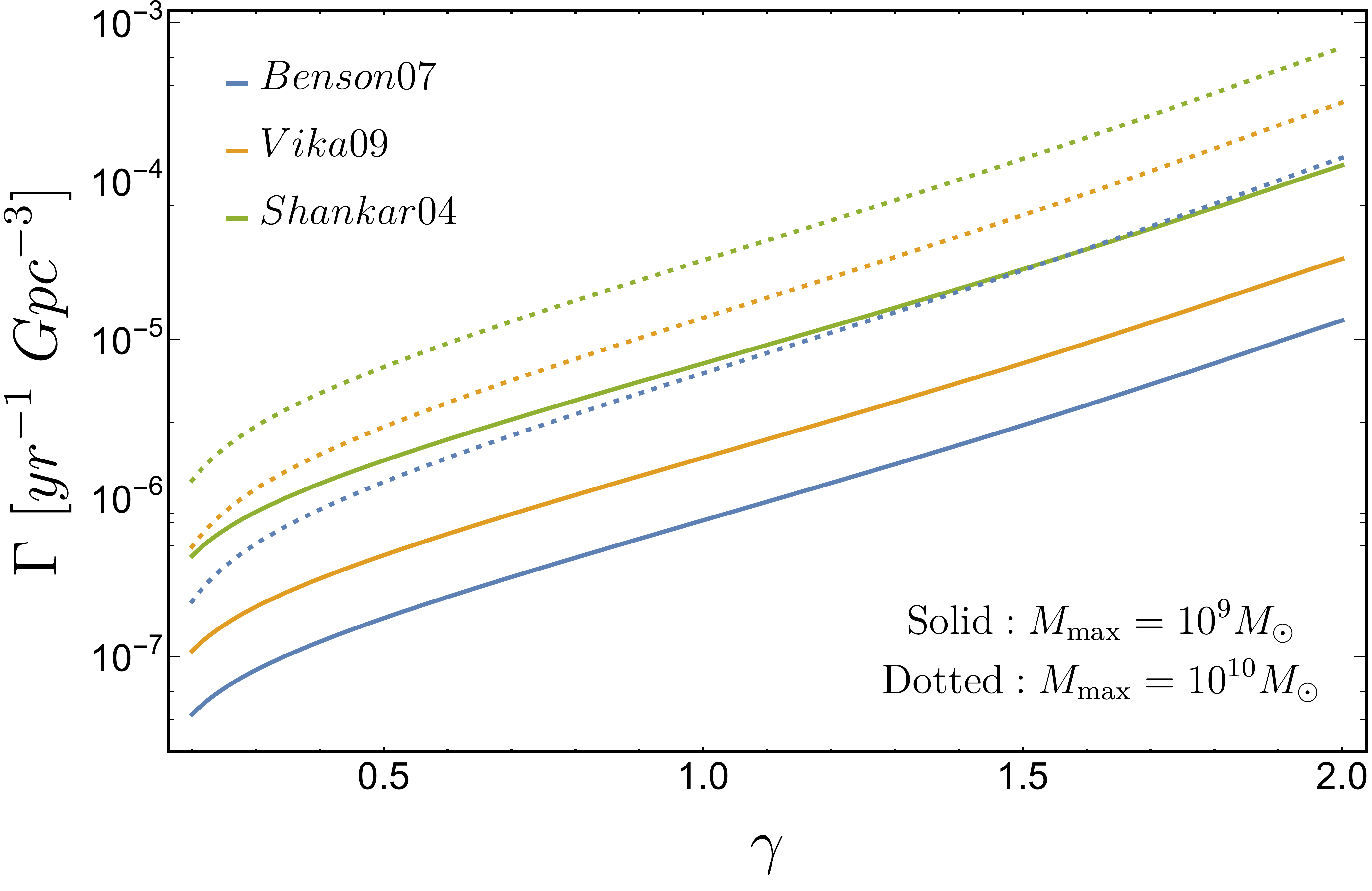}
        \caption{$\Gamma(\gamma)$ from the (fully) relaxed DM spikes with 3 different halo mass functions represented in the same colors.
        This result assumes the effect of loss-cone refilling on the DM spikes is negligible compared to the effect of relaxation.
        We plot $\Gamma$ with $M_{\max}=10^9 \, M_{\odot}$ as solid lines and $M_{\max}=10^{10} \, M_{\odot}$ as dotted lines.
        Compared to the rate $\Gamma$ from the original DM spikes in Fig.~\ref{fig:ratevsgamma},  it is significantly reduced for all $\gamma$ values.}\label{fig:raterelvsgamma}
\end{figure}

In the relaxed-spike limit, the smoothed DM spikes produce a distinct total rate $\Gamma(\gamma)$, as shown in Fig.~\ref{fig:raterelvsgamma}.
While the contribution to the merger rate from the outer part of the DM spike increases as the excess mass is distributed outward (see Sec.~\ref{sec:Evolution}), the contribution from the vicinity of the central SMBH ($4\Rs<r<R_{\mathrm{core}}$) is suppressed due to the formation of a relaxed core, resulting in a significant reduction of the total GW emissions from the DM-spike regions.
We plot the results with two different $M_{\max}$ values while $M_{\min}$ is fixed to $10^6 M_{\odot}$, since the upper bound $M_{\max}$ is found to introduce more uncertainty than the lower bound $M_{\min}$.
Figure~\ref{fig:raterelvsgamma} also shows that the suppression of the final rate $\Gamma(\gamma)$ due to relaxation becomes slightly more significant for the DM spikes with larger $\gamma$, compared to Fig.~\ref{fig:ratevsgamma}.
In this limit, we conclude that $\Gamma$ is lowered down to $< 1\%$ of the full rate inferred by LIGO, setting the lower bound for the GW-detection rate we predict.

Finally, we consider other factors that may further influence the final rate $\Gamma$.
In the relaxed-spike limit, we did not exclude the contribution to $\Gamma$ from PBHs that would have been merged with or slingshot away due to close encounters with the central SMBH, as the formation of a core significantly lowers the DM density in its close vicinity.
We checked that the fraction of PBHs in the DM-spike regions that have merged after $\sim 10 \, \rm Gyr$ stays negligible entirely for $0<\gamma<2$ in the relaxed DM spikes (and for $0<\gamma<1.5$ in the original DM spikes).
Lastly, we also checked the final rate $\Gamma$ in the case of DM comprised of PBHs with $M_{\mathrm{PBH}} \neq 30 M_{\odot}$: while lower mass PBHs with $20 M_{\odot} \leq M_{\mathrm{PBH}} \leq 30 M_{\odot}$ lead to less significant smoothing of the DM spike (see Eq.~\eqref{eq:relaxation}) and enhance $\Gamma$ in the relaxed-spike limit (by $\sim 10$ for $M_{\mathrm{PBH}} = 20 M_{\odot}$), it stays less than $\sim 1 \%$ of the inferred LIGO-detection rate (higher mass PBHs with $30 M_{\odot} \leq M_{\mathrm{PBH}} \leq 100 M_{\odot}$ lead to more suppression of the rate $\Gamma$ as the local relaxation time decreases).
In the original-spike limit, these considerations also require a closer examination of the PBH-replenishment mechanism into the DM spikes.
They are thus ignored in the final rate $\Gamma$ in the original-spike limit, and we present $\Gamma$ in Fig.~\ref{fig:ratevsgamma} as the upper bound, as we discussed in Sec.~\ref{sec:Evolution}.

\section{Conclusions}\label{sec:Conclusions}
The conclusion of our work is that PBH mergers occurring in the DM spikes around SMBHs at the center of DM halos may emit GWs frequently enough to significantly contribute to the total GW detection rate.
As we emphasize throughout, our incomplete knowledge of the dark-matter distribution near SMBHs (and the abundance of SMBHs in the Universe) renders our results quite uncertain.
The effect on the DM spikes of PBH replenishment though loss-cone filling, among the factors we understand poorly, merits a closer examination as it may influence the final rate significantly (see Sec.~\ref{sec:Evolution}).
The GW-detection rate we predict thus lies between the two limits we considered, each corresponding to Fig.~\ref{fig:ratevsgamma} and Fig.~\ref{fig:raterelvsgamma} respectively.
Future advances in our understanding of these quantities will allow a more precise determination of the GW-detection rate from the innermost parts of halos, following the prescription we have presented.

In our calculation, we have not considered PBH binary formation around SMBH spikes in dwarf galaxies, nor via three-body PBH binary formation rates, which are likely to be dominant over dissipative capture by GW emission in dense stellar systems~\cite{Park:2017zgj}.
These contributions to PBH binary formation can only increase our predicted rates, but will predominantly result in ejected PBH binaries.

Nevertheless, this model does lend itself to scrutiny with future GW measurements. Upcoming observations may enable us to constrain a PBH contribution to the overall mass spectrum of merging binaries~\cite{Kovetz:2016kpi,Kovetz:2017rvv}, while the measured eccentricities of the events could  be used to test the two-body formation channel~\cite{Cholis:2016kqi}.
Finally, if the actual detection rate nears the rate from the original DM spikes (Fig.~\ref{fig:ratevsgamma}), it would imply that we expect a substantial amount of GW signals to be coming from the centers of galaxies, as opposed to the smallest DM halos which do not host galaxies, under the late-Universe PBH-merger scenario~\cite{Bird:2016dcv} (not the BH binary formation scenario in the early Universe~\cite{Nakamura:1997sm}).
A future network of sensitive GW detectors with the ability to localize the detected events down to $<1^{\circ}$ can potentially be used to probe this scenario by cross-correlating maps of GW events with galaxy catalogues~\cite{Raccanelli:2016cud, Raccanelli:2016fmc}.
In contrast, if the rate is closer to the significantly lower values calculated for the relaxed DM spikes (Fig.~\ref{fig:ratevsgamma}), we would see such localizations to have little correlation with luminous-galaxy positions, which is in agreement with Ref.~\cite{Bird:2016dcv}.

\acknowledgments{}
This work was supported by NSF Grant No. 0244990, NASA NNX15AB18G, and the Simons Foundation.


\begin{thebibliography}{95}

 \bibitem{Abbott:2016blz}
  B.~P.~Abbott {\it et al.} [LIGO Scientific and Virgo Collaborations],
  ``Observation of Gravitational Waves from a Binary Black Hole Merger,''
  Phys.\ Rev.\ Lett.\  {\bf 116}, 061102 (2016)
  [arXiv:1602.03837 [gr-qc]].
  
 \bibitem{Abbott:2016nmj} 
  B.~P.~Abbott {\it et al.} [LIGO Scientific and Virgo Collaborations],
  ``GW151226: Observation of Gravitational Waves from a 22-Solar-Mass Binary Black Hole Coalescence,''
  Phys.\ Rev.\ Lett.\  {\bf 116}, 241103 (2016)
  [arXiv:1606.04855 [gr-qc]].
   
\bibitem{Abbott:2017vtc} 
  B.~P.~Abbott {\it et al.} [LIGO Scientific and VIRGO Collaborations],
  Phys.\ Rev.\ Lett.\  {\bf 118}, no. 22, 221101 (2017)
  Erratum: [Phys.\ Rev.\ Lett.\  {\bf 121}, no. 12, 129901 (2018)]
  [arXiv:1706.01812 [gr-qc]].

\bibitem{Belczynski:2009xy} 
  K.~Belczynski, T.~Bulik, C.~L.~Fryer, A.~Ruiter, J.~S.~Vink and J.~R.~Hurley,
  ``On The Maximum Mass of Stellar Black Holes,''
  Astrophys.\ J.\  {\bf 714}, 1217 (2010)
  [arXiv:0904.2784 [astro-ph.SR]].

\bibitem{Belczynski:2016obo} 
  K.~Belczynski, D.~E.~Holz, T.~Bulik and R.~O'Shaughnessy,
  ``The first gravitational-wave source from the isolated evolution of two 40-100 Msun stars,''
  Nature {\bf 534}, 512 (2016)
  [arXiv:1602.04531 [astro-ph.HE]].

 \bibitem{TheLIGOScientific:2016pea} 
  B.~P.~Abbott {\it et al.} [LIGO Scientific and Virgo Collaborations],
  ``Binary Black Hole Mergers in the First Advanced LIGO Observing Run,''
  Phys.\ Rev.\ X {\bf 6}, 041015 (2016)
  [arXiv:1606.04856 [gr-qc]].

 \bibitem{Kovetz:2016kpi} 
  E.~D.~Kovetz, I.~Cholis, P.~C.~Breysse and M.~Kamionkowski,
  ``Black hole mass function from gravitational wave measurements,''
  Phys.\ Rev.\ D {\bf 95}, 103010 (2017)
  [arXiv:1611.01157 [astro-ph.CO]].

 \bibitem{Zel'dovich:1967}
  Zel'dovich, Ya B., and I. D. Novikov,
  ``The Hypothesis of Cores Retarded during Expansion and the Hot Cosmological Model,''
  Sov.\ Astron.\ {\bf 10}, 602 (1967).

 \bibitem{Carr:1974nx} 
  B.~J.~Carr and S.~W.~Hawking,
  ``Black holes in the early Universe,''
  Mon.\ Not.\ Roy.\ Astron.\ Soc.\  {\bf 168}, 399 (1974).

 \bibitem{Chapline:1975} 
  G.~F.~Chapline,
  ``Cosmological effects of primordial black holes,''
  Nature {\bf 253}, 251 (1975)

 \bibitem{Frampton:2010sw} 
  P.~H.~Frampton, M.~Kawasaki, F.~Takahashi and T.~T.~Yanagida,
  ``Primordial Black Holes as All Dark Matter,''
  JCAP {\bf 1004}, 023 (2010)
  [arXiv:1001.2308 [hep-ph]].
  
 \bibitem{Bird:2016dcv} 
  S.~Bird, I.~Cholis, J.~B.~Mu\~noz, Y.~Ali-Ha\"{i}moud, 
  M.~Kamionkowski, E.~D.~Kovetz, A.~Raccanelli and A.~G.~Riess,
  ``Did LIGO detect dark matter?,''
  Phys.\ Rev.\ Lett.\  {\bf 116}, 20, 201301 (2016)
  [arXiv:1603.00464 [astro-ph.CO]].
  
 \bibitem{Clesse:2016vqa} 
  S.~Clesse and J.~Garc\'\i a-Bellido,
  ``The clustering of massive Primordial Black Holes as Dark Matter: measuring their mass distribution with Advanced LIGO,''
  Phys.\ Dark Univ.\  {\bf 15}, 142 (2017)
  [arXiv:1603.05234 [astro-ph.CO]].
  
 \bibitem{Sasaki:2016jop} 
  M.~Sasaki, T.~Suyama, T.~Tanaka and S.~Yokoyama,
  ``Primordial Black Hole Scenario for the Gravitational-Wave Event GW150914,''
  Phys.\ Rev.\ Lett.\  {\bf 117}, 061101 (2016)
  [arXiv:1603.08338].

 \bibitem{Kashlinsky:2016sdv} 
  A.~Kashlinsky,
  ``LIGO gravitational wave detection, primordial black holes and the near-IR cosmic infrared background anisotropies,''
  Astrophys.\ J.\  {\bf 823}, L25 (2016)
  [arXiv:1605.04023 [astro-ph.CO]].

 \bibitem{Blinnikov:2016bxu} 
  S.~Blinnikov, A.~Dolgov, N.~K.~Porayko and K.~Postnov,
  ``Solving puzzles of GW150914 by primordial black holes,''
  JCAP {\bf 1611}, 036 (2016)
  [arXiv:1611.00541 [astro-ph.HE]].

 \bibitem{Dolgov:1992pu} 
  A.~Dolgov and J.~Silk,
  ``Baryon isocurvature fluctuations at small scales and baryonic dark matter,''
  Phys.\ Rev.\ D {\bf 47}, 4244 (1993).

 \bibitem{Dolgov:2008wu} 
  A.~D.~Dolgov, M.~Kawasaki and N.~Kevlishvili,
  ``Inhomogeneous baryogenesis, cosmic antimatter, and dark matter,''
  Nucl.\ Phys.\ B {\bf 807}, 229 (2009)
  [arXiv:0806.2986 [hep-ph]].

 \bibitem{Alcock:1996yv} 
  C.~Alcock {\it et al.} [MACHO Collaboration],
  ``The MACHO project LMC microlensing results from the first two years and the nature of the galactic dark halo,''
  Astrophys.\ J.\  {\bf 486}, 697 (1997)
  [astro-ph/9606165].
  
 \bibitem{Tisserand:2006zx} 
  P.~Tisserand {\it et al.} [EROS-2 Collaboration],
  ``Limits on the Macho Content of the Galactic Halo from the EROS-2 Survey of the Magellanic Clouds,''
  Astron.\ Astrophys.\  {\bf 469}, 387 (2007)
  [astro-ph/0607207].
  
 \bibitem{Mediavilla:2017bok} 
  E.~Mediavilla, J.~Jimenez-Vicente, J.~A.~Mu\~noz, H.~Vives-Arias and J.~Calderon-Infante,
  ``Limits on the Mass and Abundance of Primordial Black Holes from Quasar Gravitational Microlensing,''
  Astrophys.\ J.\  {\bf 836}, L18 (2017)
  [arXiv:1702.00947].

 \bibitem{Ricotti:2007au} 
  M.~Ricotti, J.~P.~Ostriker, and K.~J.~Mack,
  ``Effect of Primordial Black Holes on the Cosmic Microwave Background and Cosmological Parameter Estimates,''
  Astrophys.\ J.\  {\bf 680}, 829 (2008)
  [arXiv:0709.0524 [astro-ph]].
  
 \bibitem{Ali-Haimoud:2016mbv} 
  Y.~Ali-Ha\"{i}moud and M.~Kamionkowski,
  ``Cosmic microwave background limits on accreting primordial black holes,''
  Phys.\ Rev.\ D {\bf 95}, 043534 (2017)
  [arXiv:1612.05644].
      
 \bibitem{Brandt:2016aco} 
  T.~D.~Brandt,
  ``Constraints on MACHO Dark Matter from Compact Stellar Systems in Ultra-Faint Dwarf Galaxies,''
  Astrophys.\ J.\  {\bf 824}, L31 (2016)
  [arXiv:1605.03665].
  
 \bibitem{Raidal:2017mfl} 
  M.~Raidal, V.~Vaskonen and H.~Veerm{\"a}e,
  ``Gravitational Waves from Primordial Black Hole Mergers,''
  arXiv:1707.01480 [astro-ph.CO].
  
\bibitem{Poulin:2017bwe} 
  V.~Poulin, P.~D.~Serpico, F.~Calore, S.~Clesse and K.~Kohri,
  ``CMB bounds on disk-accreting massive primordial black holes,''
  Phys.\ Rev.\ D {\bf 96}, no. 8, 083524 (2017)
  [arXiv:1707.04206 [astro-ph.CO]].

 \bibitem{Munoz:2016tmg} 
  J.~B.~Mu\~noz, E.~D.~Kovetz, L.~Dai and M.~Kamionkowski,
  ``Lensing of Fast Radio Bursts as a Probe of Compact Dark Matter,''
  Phys.\ Rev.\ Lett.\  {\bf 117}, 091301 (2016)
  [arXiv:1605.00008].
  
 \bibitem{Schutz:2016khr} 
  K.~Schutz and A.~Liu,
  ``Pulsar timing can constrain primordial black holes in the LIGO mass window,''
  Phys.\ Rev.\ D {\bf 95}, 023002 (2017)
  [arXiv:1610.04234].

\bibitem{Clesse:2016ajp} 
  S.~Clesse and J.~García-Bellido,
  ``Detecting the gravitational wave background from primordial black hole dark matter,''
  Phys.\ Dark Univ.\  {\bf 18}, 105 (2017)
  [arXiv:1610.08479 [astro-ph.CO]].

\bibitem{Kovetz:2017rvv} 
  E.~D.~Kovetz,
  ``Probing Primordial-Black-Hole Dark Matter with Gravitational Waves,''
  Phys.\ Rev.\ Lett.\  {\bf 119}, no. 13, 131301 (2017)
  [arXiv:1705.09182 [astro-ph.CO]].

\bibitem{Venumadhav:2017pps} 
  T.~Venumadhav, L.~Dai and J.~Miralda-Escudé,
  ``Microlensing of Extremely Magnified Stars near Caustics of Galaxy Clusters,''
  Astrophys.\ J.\  {\bf 850}, no. 1, 49 (2017)
  [arXiv:1707.00003 [astro-ph.CO]].

\bibitem{Diego:2017drh} 
  J.~M.~Diego {\it et al.},
  ``Dark Matter under the Microscope: Constraining Compact Dark Matter with Caustic Crossing Events,''
  Astrophys.\ J.\  {\bf 857}, no. 1, 25 (2018)
  [arXiv:1706.10281 [astro-ph.CO]].

 \bibitem{Hayasaki:2009ug} 
  K.~Hayasaki, K.~Takahashi, Y.~Sendouda and S.~Nagataki,
  ``Rapid merger of binary primordial black holes: An implication for GW150914,''
  Publ.\ Astron.\ Soc.\ Jap.\  {\bf 68}, 66 (2016)
  [arXiv:0909.1738 [astro-ph.CO]].

\bibitem{Ali-Haimoud:2017rtz} 
  Y.~Ali-Ha\"{i}moud, E.~D.~Kovetz and M.~Kamionkowski,
  ``Merger rate of primordial black-hole binaries,''
  Phys.\ Rev.\ D {\bf 96}, no. 12, 123523 (2017)
  [arXiv:1709.06576 [astro-ph.CO]].

 \bibitem{Gondolo:1999ef} 
  P.~Gondolo and J.~Silk,
  ``Dark matter annihilation at the galactic center,''
  Phys.\ Rev.\ Lett.\  {\bf 83}, 1719 (1999)
  [astro-ph/9906391].
 
 \bibitem{Graham:2005xx} 
  A.~W.~Graham, D.~Merritt, B.~Moore, J.~Diemand and B.~Terzic,
  ``Empirical models for Dark Matter Halos. I. Nonparametric Construction of Density Profiles and Comparison with Parametric Models,''
  Astron.\ J.\  {\bf 132}, 2685 (2006)
  [astro-ph/0509417].
  
 \bibitem{Stadel:2008pn} 
  J.~Stadel, D.~Potter, B.~Moore, J.~Diemand, P.~Madau, M.~Zemp, M.~Kuhlen and V.~Quilis,
  ``Quantifying the heart of darkness with GHALO - a multi-billion particle simulation of our galactic halo,''
  Mon.\ Not.\ Roy.\ Astron.\ Soc.\  {\bf 398}, L21 (2009)
  [arXiv:0808.2981 [astro-ph]].
  
 \bibitem{Navarro:2008kc} 
  J.~F.~Navarro {\it et al.},
  ``The Diversity and Similarity of Cold Dark Matter Halos,''
  Mon.\ Not.\ Roy.\ Astron.\ Soc.\  NRAS  {\bf 402}, 21 (2010)
  [arXiv:0810.1522 [astro-ph]].

 \bibitem{Ullio:2001fb} 
  P.~Ullio, H.~S.~Zhao and M.~Kamionkowski,
  ``A Dark matter spike at the galactic center?,''
  Phys.\ Rev.\ D {\bf 64}, 043504 (2001)
  [astro-ph/0101481].

\bibitem{Ferrer:2017xwm} 
  F.~Ferrer, A.~Medeiros da Rosa and C.~M.~Will,
  ``Dark matter spikes in the vicinity of Kerr black holes,''
  Phys.\ Rev.\ D {\bf 96}, no. 8, 083014 (2017)
  [arXiv:1707.06302 [astro-ph.CO]].

 \bibitem{Ferrarese:2000se} 
  L.~Ferrarese and D.~Merritt,
  ``A Fundamental relation between supermassive black holes and their host galaxies,''
  Astrophys.\ J.\  {\bf 539}, L9 (2000)
  [astro-ph/0006053].
  
 \bibitem{Gebhardt:2000fk} 
  K.~Gebhardt {\it et al.},
  ``A Relationship between nuclear black hole mass and galaxy velocity dispersion,''
  Astrophys.\ J.\  {\bf 539}, L13 (2000)
  [astro-ph/0006289].

 \bibitem{Kormendy:1995er} 
  J.~Kormendy and D.~Richstone,
  ``Inward bound: The Search for supermassive black holes in galactic nuclei,''
  Ann.\ Rev.\ Astron.\ Astrophys.\  {\bf 33}, 581 (1995).

 \bibitem{Magorrian:1997hw} 
  J.~Magorrian {\it et al.},
  ``The Demography of massive dark objects in galaxy centers,''
  Astron.\ J.\  {\bf 115}, 2285 (1998)
  [astro-ph/9708072].

 \bibitem{Marconi:2003hj} 
  A.~Marconi and L.~K.~Hunt,
  ``The relation between black hole mass, bulge mass, and near-infrared luminosity,''
  Astrophys.\ J.\  {\bf 589}, L21 (2003)
  [astro-ph/0304274].
 
 \bibitem{Navarro:1995iw} 
  J.~F.~Navarro, C.~S.~Frenk and S.~D.~M.~White,
  ``The Structure of cold dark matter halos,''
  Astrophys.\ J.\  {\bf 462}, 563 (1996)
  [astro-ph/9508025].

 \bibitem{Gultekin:2009qn} 
  K.~Gultekin {\it et al.},
  ``The M-sigma and M-L Relations in Galactic Bulges and Determinations of their Intrinsic Scatter,''
  Astrophys.\ J.\  {\bf 698}, 198 (2009)
  [arXiv:0903.4897 [astro-ph.GA]].
  
 \bibitem{Kormendy:2013dxa} 
  J.~Kormendy and L.~C.~Ho,
  ``Coevolution (Or Not) of Supermassive Black Holes and Host Galaxies,''
  Ann.\ Rev.\ Astron.\ Astrophys.\  {\bf 51}, 511 (2013)
  [arXiv:1304.7762 [astro-ph.CO]].
  
 \bibitem{Prada:2011jf} 
  F.~Prada, A.~A.~Klypin, A.~J.~Cuesta, J.~E.~Betancort-Rijo and J.~Primack,
  ``Halo concentrations in the standard LCDM cosmology,''
  Mon.\ Not.\ Roy.\ Astron.\ Soc.\  {\bf 423}, 3018 (2012)
  [arXiv:1104.5130 [astro-ph.CO]].
  
 \bibitem{Dzanovic:2006px} 
  A.~J.~Benson, D.~Dzanovic, C.~S.~Frenk and R.~Sharples,
  ``Luminosity and stellar mass functions of disks and spheroids in the SDSS and the supermassive black hole mass function,''
  Mon.\ Not.\ Roy.\ Astron.\ Soc.\  {\bf 379}, 841 (2007)
  [astro-ph/0612719].
  
 \bibitem{Vika:2009ef} 
  M.~Vika, S.~P.~Driver, A.~W.~Graham and J.~Liske,
  ``The Millennium Galaxy Catalogue: The $M_{bh}$--$L_{spheroid}$ derived supermassive black hole mass function,''
  Mon.\ Not.\ Roy.\ Astron.\ Soc.\  {\bf 400}, 1451 (2009)
  [arXiv:0908.2102 [astro-ph.CO]].
   
 \bibitem{Liske:2002nk} 
  J.~Liske, D.~J.~Lemon, S.~P.~Driver, N.~J.~G.~Cross and W.~J.~Couch,
  ``The millennium galaxy catalogue: $16 < b_{MGC} < 24$ Galaxy counts and the calibration of the local galaxy luminosity function,''
  Mon.\ Not.\ Roy.\ Astron.\ Soc.\  {\bf 344}, 307 (2003)
  [astro-ph/0207555].
  
 \bibitem{Shankar:2004ir} 
  F.~Shankar, P.~Salucci, G.~L.~Granato, G.~De Zotti and L.~Danese,
  ``Super-massive black hole demography: The Match between the local and accreted mass functions,''
  Mon.\ Not.\ Roy.\ Astron.\ Soc.\  {\bf 354}, 1020 (2004)
  [astro-ph/0405585].
 
 \bibitem{Baldassare:2016cox} 
  V.~F.~Baldassare, A.~E.~Reines, E.~Gallo and J.~E.~Greene,
  ``X-ray and Ultraviolet Properties of AGN in Nearby Dwarf Galaxies,''
  Astrophys.\ J.\  {\bf 836}, 20 (2017)
  [arXiv:1609.07148 [astro-ph.HE]].

 \bibitem{Silk:2017yai} 
  J.~Silk,
  ``Feedback by Massive Black Holes in Gas-rich Dwarf Galaxies,''
  Astrophys.\ J.\  {\bf 839}, no. 1, L13 (2017)
  [arXiv:1703.08553 [astro-ph.GA]].

\bibitem{Sellwood:2010qa} 
  J.~A.~Sellwood,
  ``Dynamics of Disks and Warps,''
  In: Oswalt T.D., Gilmore G. (eds) Planets, Stars and Stellar Systems. Springer, Dordrecht (2013)
  [arXiv:1006.4855 [astro-ph.GA]].

\bibitem{Lightman:1977zz} 
  A.~P.~Lightman and S.~L.~Shapiro,
  ``The distribution and consumption rate of stars around a massive, collapsed object,''
  Astrophys.\ J.\  {\bf 211}, 244 (1977).

\bibitem{Frank:1976uy} 
  J.~Frank and M.~J.~Rees,
  ``Effects of massive central black holes on dense stellar systems,''
  Mon.\ Not.\ Roy.\ Astron.\ Soc.\  {\bf 176}, 633 (1976).

\bibitem{Merritt:2013awa} 
  D.~Merritt,
  ``Loss-cone Dynamics,''
  Class.\ Quant.\ Grav.\  {\bf 30}, 244005 (2013)
  [arXiv:1307.3268 [astro-ph.GA]].

\bibitem{Vasiliev:2013az} 
  E.~Vasiliev and D.~Merritt,
  ``The loss cone problem in axisymmetric nuclei,''
  Astrophys.\ J.\  {\bf 774}, 87 (2013)
  [arXiv:1301.3150 [astro-ph.GA]].

\bibitem{2017MNRAS.464.2301G} 
  A.~Gualandris, J.~I.~Read, W.~Dehnen and E.~Bortolas,
  ``Collisionless loss-cone refilling: there is no final parsec problem,''
  Mon.\ Not.\ Roy.\ Astron.\ Soc.\  {\bf 464}, 2301 (2017)
  [arXiv:1609.09383 [astro-ph.GA]].

\bibitem{Naoz:2014zda} 
  S.~Naoz and J.~Silk,
  ``Formation of Dark Matter Torii Around Supermassive Black Holes Via The Eccentric Kozai-Lidov Mechanism,''
  Astrophys.\ J.\  {\bf 795}, 102 (2014)
  [arXiv:1409.5432 [astro-ph.GA]].

\bibitem{Petrovich:2017otm} 
  C.~Petrovich and F.~Antonini,
  ``Greatly enhanced merger rates of compact-object binaries in non-spherical nuclear star clusters,''
  Astrophys.\ J.\  {\bf 846}, no. 2, 146 (2017)
  [arXiv:1705.05848 [astro-ph.HE]].

\bibitem{Sigurdsson:2003ei} 
  S.~Sigurdsson,
  ``Loss cone: past, present and future,''
  Class.\ Quant.\ Grav.\  {\bf 20}, S45 (2003)
  [astro-ph/0304251].

\bibitem{Kuhnel:2018mlr} 
  F.~Kuhnel, A.~Matas, G.~D.~Starkman and K.~Freese,
  ``Waves from the Centre: Probing PBH and other Macroscopic Dark Matter with LISA,''
  arXiv:1811.06387 [gr-qc].

 \bibitem{Park:2017zgj} 
  D.~Park, C.~Kim, H.~M.~Lee, Y.~B.~Bae and C.~Belczynski,
  ``Black Hole Binaries Dynamically Formed in Globular Clusters,''
  Mon.\ Not.\ Roy.\ Astron.\ Soc.\  {\bf 469}, 4665 (2017)
  [arXiv:1703.01568 [astro-ph.HE]].

 \bibitem{Cholis:2016kqi} 
  I.~Cholis, E.~D.~Kovetz, Y.~Ali-Ha\"{i}moud, S.~Bird, M.~Kamionkowski, J.~B.~Mu\~noz and A.~Raccanelli,
  ``Orbital eccentricities in primordial black hole binaries,''
  Phys.\ Rev.\ D {\bf 94}, 084013 (2016)
  [arXiv:1606.07437].

\bibitem{Nakamura:1997sm} 
  T.~Nakamura, M.~Sasaki, T.~Tanaka and K.~S.~Thorne,
  ``Gravitational waves from coalescing black hole MACHO binaries,''
  Astrophys.\ J.\  {\bf 487}, L139 (1997)
  [astro-ph/9708060].
 \bibitem{Raccanelli:2016cud} 
  A.~Raccanelli, E.~D.~Kovetz, S.~Bird, I.~Cholis and J.~B.~Mu\~noz,
  ``Determining the progenitors of merging black-hole binaries,''
  Phys.\ Rev.\ D {\bf 94}, 023516 (2016)
  [arXiv:1605.01405 [astro-ph.CO]].

 \bibitem{Raccanelli:2016fmc} 
  A.~Raccanelli,
  ``Gravitational wave astronomy with radio galaxy surveys,''
  Mon.\ Not.\ Roy.\ Astron.\ Soc.\  {\bf 469}, 656 (2017)
  [arXiv:1609.09377 [astro-ph.CO]].

\end{thebibliography}
\end{document}